\documentclass{PoS}

\usepackage{graphicx}

\def\OMIT#1{{}}

\def\vereq#1#2{\lower3.5pt\vbox{\baselineskip1.5pt \lineskip1.5pt
\ialign{$#1\hfill##\hfil$\crcr#2\crcr\sim\crcr}}}

\def\d{{\rm d}}
\def\ov{\overline}

\def\TeV{{\rm TeV}}

\def\fbm{\ensuremath{{\rm fb}^{-1}}}
\def\pbm{\ensuremath{{\rm pb}^{-1}}}

\newcommand{\beq}{\begin{equation}}
\newcommand{\eeq}{\end{equation}}
\newcommand{\beqa}{\begin{eqnarray}}
\newcommand{\eeqa}{\end{eqnarray}}

\newcommand\url[1]{{\tt\href{#1}{#1}}}

\title{Supermodels: Early new physics at the LHC?}

\ShortTitle{Supermodels: Early new physics at the LHC?}

\author{\speaker{Zoltan Ligeti}\\
Lawrence Berkeley National Laboratory,
University of California, Berkeley, CA 94720, USA\\
E-mail: \email{ligeti@lbl.gov}}

\abstract{We investigate new physics that could be discovered with very little
LHC data, beyond the expected sensitivity of the Tevatron.  We construct
``supermodels", for which the LHC sensitivity with 10~pb$^{-1}$ luminosity is
already greater than that of the Tevatron with 10~fb$^{-1}$.  The simplest
supermodels involve $s$-channel resonances in the quark-antiquark and especially
in the quark-quark channels.  In the latter case, the LHC sensitivity with
0.1~pb$^{-1}$ can already be greater than that of the Tevatron with
10~fb$^{-1}$.  We concentrate on easily visible final states with small standard
model backgrounds, and find that there are simple searches, besides those for
$Z'$ states, which could discover new physics in early LHC data.  Many of these
are well-suited to test searches for ``conventional" models, often discussed for
larger data sets.}

\FullConference{35th International Conference of High Energy Physics -- ICHEP2010\\
		July 22--28, 2010\\
		Paris France}

\begin{document}

\section{Introduction}

We would like to explore what is the minimal luminosity the LHC needs in order
to possibly discover new physics beyond the sensitivity of the
Tevatron~\cite{Bauer:2009cc}.  This talk concentrates on which new physics
signatures could be discovered with a few 10s of \pbm\ luminosity, beyond the
reach of the $\sim$10~\fbm\ Tevatron data expected by the end of 2010.

The definition of a \textit{supermodel\/} in the Merriam-Webster Dictionary is
``a famous and successful fashion model" --- we define it as the class of models
/ Lagrangians which the LHC can discover with small luminosity, satisfying the
following criteria:
\begin{enumerate}\vspace*{-6pt}\itemsep 0pt
\setlength{\parskip}{0pt}\setlength{\parsep}{0pt}
\item Large enough LHC cross section to produce at least 10 signal
events\footnote{While fewer events may be sufficient for discovery, we demand 10
to allow for $\mathcal{O}(1)$ uncertainties in our analysis.}
with 10~pb$^{-1}$ of data;
\item Small enough Tevatron cross section to evade the 2010 Tevatron
sensitivity with $10\ \fbm$;
\item Large enough branching fraction to an ``easy'' final state with
essentially no backgrounds;
\item Consistency with other existing bounds.
\end{enumerate}\vspace*{-6pt}
\noindent
However, we are not concerned with solving usual model building goals, such as 
unification, weak scale dark matter, or the hierarchy problem.  Thus we got some
criticism, e.g., that ``Unfortunately, the defining property of supermodels is
that they are unattainable"~\cite{Peskin}.  So it was amusing to see after this
presentation the first ATLAS~\cite{atlastalk} and CMS~\cite{cmstalk} search
results for resonances in the $qq$ channel, disproving this statement by
obtaining limits that went beyond the Tevatron bounds for the first time.  In
addition, a special presentation by N.~Sarkozy~\cite{Sarkozy} was announced soon
after this talk, who also disproved the claimed unattainability of supermodels
using different techniques.

\section{Resonance scenarios}

To determine if a new physics scenario can be a supermodel, we need to compare 
$[{\cal L} \times \sigma \times {\cal B} \times {\cal E}]$ at the LHC and the
Tevatron, where ${\cal L}$ is the luminosity, $\sigma$ is the cross section, 
${\cal B}$ is the branching ratio into a detected mode, and ${\cal E}$ is the
efficiency.  Since the Tevatron and LHC detectors are similar, to get a rough
estimate, we can take ${\cal E}_{\rm LHC} \approx {\cal E}_{\rm TEV}$.  If we
further assume that the detection modes are the same, then ${\cal B}_{\rm LHC} =
{\cal B}_{\rm TEV}$, and we simply need $\sigma_{\rm LHC} / \sigma_{\rm TEV} >
{\cal L}_{\rm TEV} / {\cal L}_{\rm LHC}$, i.e., ${\cal O}(1000)$ times larger
LHC than Tevatron cross sections.

\begin{figure}[t]
\centerline{\includegraphics[width=.5\textwidth]{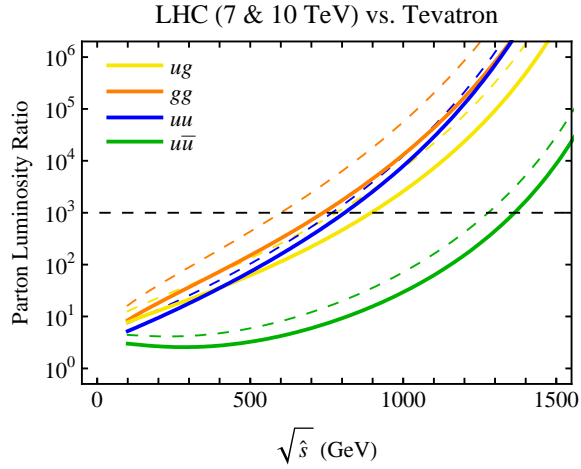}}
\caption{Ratios of the parton luminosities for 7 TeV (solid) and 10 TeV (dashed)
LHC compared to the 1.96 TeV Tevatron, as functions of the partonic invariant
mass.  The dashed horizontal line shows $10^3$.}
\label{fig:lumiratios}
\end{figure}

The cross section of any process is given by
\beq\label{eq:partonluminosity}
\frac{\d\sigma}{\d \hat s} = \sum_{ij} \hat\sigma_{ij}(\hat s)\,
  {\cal F}_{ij}(\hat s, s) \,, \qquad
{\cal F}_{ij}(\hat s, s) = \int_0^1\! \d x_i\, \d x_j\, 
  f_i(x_i)\, f_j(x_j)\, \delta\big(1 - x_i\, x_j\, s/\hat s\big)\,,
\eeq
where $i,j$ denote initial partons, $\hat s$ is the partonic center-of-mass
energy while $s$ is that of the collider, $f_i(x_i)$ denote the parton
distribution functions (PDFs), and ${\cal F}_{ij}(\hat s, s)$ is called the
parton luminosity.  If one partonic $ij$ channel and a narrow $\hat s$ range
dominate, then $\sigma_{\rm LHC}/\sigma_{\rm TEV} \simeq {\cal F}_{ij}(s_{\rm
LHC}, \hat s)/{\cal F}_{ij}(2\,\TeV, \hat s)$.  The dominance of a narow $\hat
s$ range is quite generic, and is only a mild assumption for a first discovery,
because the PDFs are steeply falling in the region relevant when only a few
events are produced yet.  We plot in Figure~\ref{fig:lumiratios} the ratios of
parton luminosities at the  LHC and the Tevatron.  We used the CTEQ-5L
PDFs~\cite{Lai:1999wy} implemented in Mathematica and checked that MSTW
2008~\cite{Martin:2009iq} gives compatible results at the level of accuracy we
require.  We see that for large enough $\hat s$, the parton luminosities at the
LHC are indeed $>\!1000$ times larger than at the Tevatron.

To identify possible supermodel scenarios, we need to consider, besides the
parton luminosities, the production rates and the decay rates to visible
channels.  In general, smaller production rates and smaller detectable branching
ratios favor the Tevatron, because requiring enough events at the LHC for a
discovery reduces the $\hat s$ values probed, thereby reducing the LHC's
advantage.

Figure~\ref{fig:lumiratios} shows that at moderate values of $\hat s$, the $gg$
parton luminosity is the most enhanced (hence the LHC is often called a gluon
collider).  Then the simplest process is the pair production of new colored
particles, $gg\to X\ov X$, which can indeed have ${\cal O}({\rm pb})$ cross
sections at the LHC, promising ${\cal O}(10)$ events with ${\cal O}(10\,\pbm)$
data.  However, even assuming that $X$ and $\ov X$ decay to highly visible final
states, the LHC's sensitivity only surpasses the Tevatron's with ${\cal
O}(50\,\pbm)$ luminosity~\cite{Bauer:2009cc}.  The main reason is that the same
final state can also be produced from $q\bar q$ initial states, where the LHC's
advantage is less.  Since  QCD pair production is well-studied in many specific
new physics scenarios, and the advantage of the early LHC over the Tevatron can
only be marginal, we do not consider it to be a supermodel.

The production of an $s$-channel resonance has the potential to be a supermodel
if it has a large coupling to the partonic initial state, since the production
cross section for a single resonance is enhanced over pair production by a phase
space factor, $16 \pi^2$.  A resonance can have ${\cal O}(1)$ dimension-4
couplings to $q\bar q$ and $qq$ initial states.  However, for the $qg$ or $gg$
initial states, $SU(3)$ gauge invariance forbids renormalizable couplings to a
single resonance.  The lowest dimension operator for the $gg$ initial state is a
dimension-5 operator, $[g_s^2/(16 \pi^2 \Lambda)]\, X\, G_{\mu\nu} G^{\mu\nu}$. 
The coefficient has been estimated assuming perturbative physics at $\Lambda
\sim 1\,\TeV$, with the $1/(16 \pi^2)$ factor coming from a loop.  (For the $qg$
initial state, one can produce an excited quark via a coupling of the form
$[g_s^2/(16 \pi^2 \Lambda)]\, \ov q\, \sigma_{\mu\nu} G^{\mu\nu} X$, which can
also arise only from a loop diagram in a perturbative scenario.)  If there is
TeV-scale strong dynamics involving $X$, then the coefficients can be enhanced
up to their naive dimensional analysis value, $g_s^2/(4 \pi \Lambda)$.  However,
such strong dynamics near the TeV scale is constrained by precision
measurements, and we adopt the perturbative estimate $g^2_{\rm eff} \sim [1/(16
\pi^2)]^2$ for both $gg$ and $qg$ resonances.  These considerations rule out
supermodel resonances coupling dominantly to $qg$ or $gg$ initial
states~\cite{Bauer:2009cc}.  (Of course, if one allows ${\cal O}(1)$ couplings,
without the suppression factors suggested by naive dimensional analysis included
here, then there are other possible supermodel resonance
scenarios~\cite{Barbieri:2010nc}.)

\begin{figure*}[t]
\includegraphics[width=0.45\textwidth]{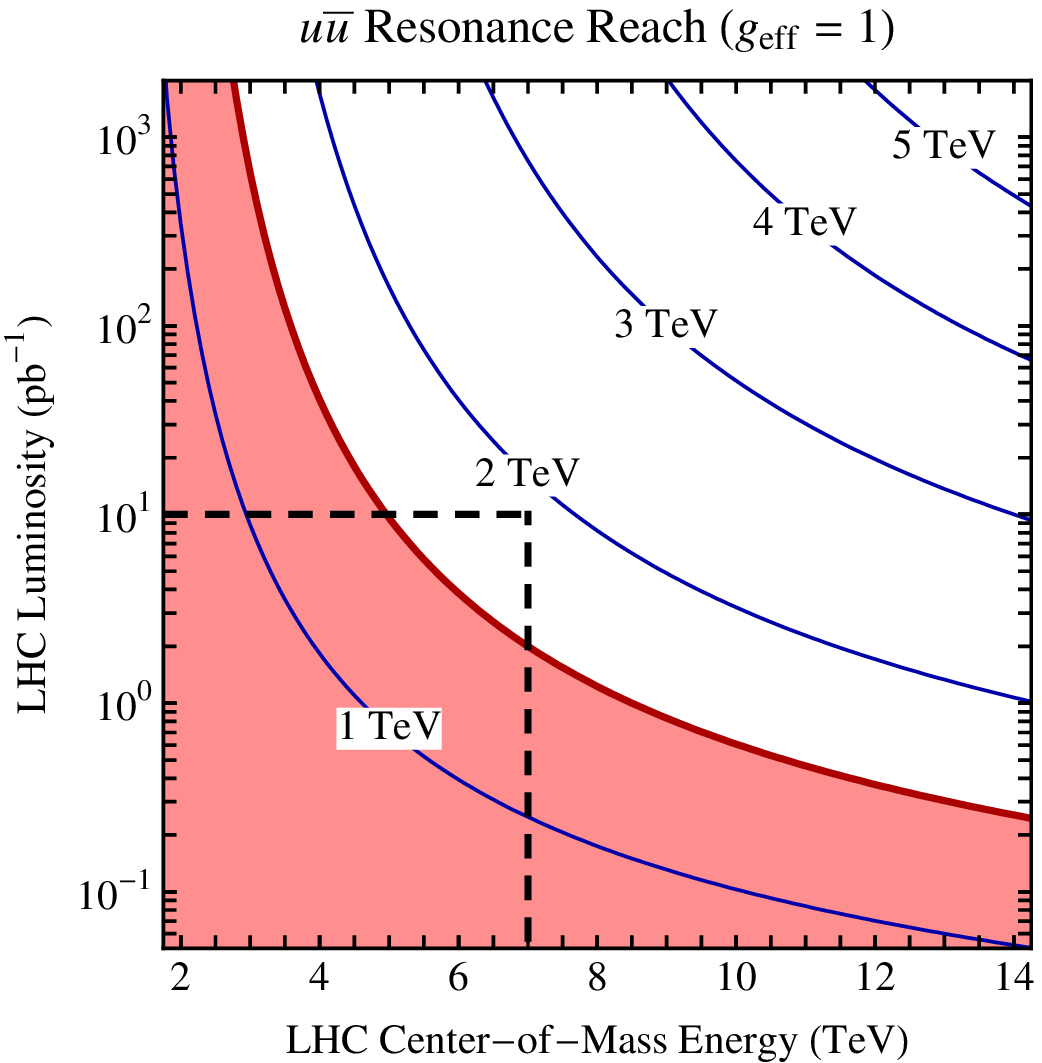}
\hfil
\includegraphics[width=0.45\textwidth]{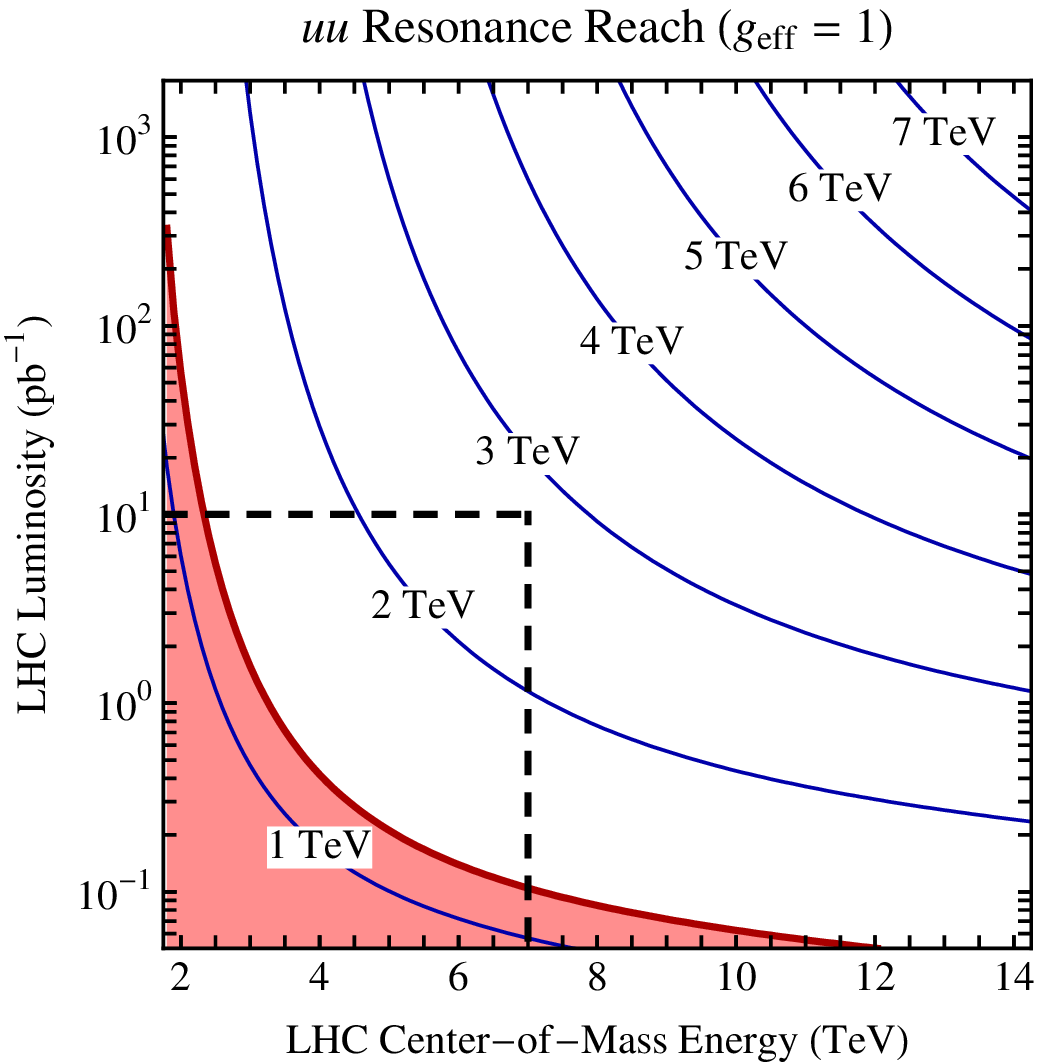}
\caption{The LHC reach for resonance production, as a function of energy and
luminosity.  The contours show the production of 10 events for a given resonance
mass.  The red regions show the Tevatron sensitivity with 10~fb$^{-1}$, and the
intersection of the dashed lines shows the maximal resonance mass probed by the
7 TeV LHC with 10~pb$^{-1}$.  The LHC exceeds the Tevatron sensitivity for $qq$
resonances already with $\sim 0.1\,\pbm$.}
\label{fig:reach}
\end{figure*}

Figure~\ref{fig:reach} shows our estimate of the generic early LHC reach in
$m_X$, as a function of the energy and luminosity, for $q\ov q$ and $qq$ 
resonances, assuming 100\% branching fraction to highly visible final states
(for concreteness, for $q=u$), showing that such resonances can be supermodels. 
For simplicity, we call any $q\ov q$ resonance a $Z'$ (even if it is a
KK-gluon), and any $qq$ resonance a ``diquark" (even if it is a colored
scalar).  Figure~\ref{fig:reach} shows that the LHC sensitivity for $q\ov q$
resonances surpasses the Tevatron with ${\cal O}(10)\,\pbm$ data.  However, for
$qq$ resonances, the LHC surpasses the Tevatron already with ${\cal
O}(0.1)\,\pbm$ at 7~TeV~\cite{Bauer:2009cc}, as demonstrated experimentally at
this Conference~\cite{atlastalk, cmstalk}.

\section{Supermodel building}

\paragraph{$q\ov q$ resonances:}

Until recently, the most often discussed new physics scenarios for early LHC
discoveries have been $Z'$ models.  However, a model with a $Z'$ coupling to
leptons and quarks is strongly constrained by the LEP~II limits on four-fermion
operators.  Flavor universal models face the problem that the production rate
is proportional to the $Z'$ coupling to quarks, $\sigma(q\ov q \to Z') \propto
g_q^2$, while the branching ratio to $\ell^+ \ell^-$ is suppressed by it, ${\cal
B}(Z' \to \ell^+ \ell^-) \propto g_\ell^2/(2 g_\ell^2 + 6 g_q^2)$.  In this
class of models there is no value of $m_{Z'}$ for which the LHC can see 10 $Z'
\to \ell^+ \ell^-$ events with $10\,\pbm$, without violating other
bounds~\cite{Bauer:2009cc}.  This conclusion can be evaded by coupling the $Z'$
only to muons, studied in detail also in Ref.~\cite{Salvioni:2009mt}.  A
$B-3L_\mu$ boson is a supermodel, but it ain't pretty.

If other particles are introduced and the $Z'$ decays with large branching
fraction to non-SM particles then many possibilities open up.  For example, a
$q\ov q$ resonance decaying to new quasi-stable charged particles can have large
branching ratios, avoid flavor physics bounds, and be cosmologically safe. 
Another option is if the resonance decays to hidden-valey type states, which
then have small couplings to decay back to SM particles; though this is unlikely
to be easily reconstructible.

\paragraph{$qq$ resonances:}

As can be seen from Figure~\ref{fig:reach}, enormous cross sections are possible
for a resonance in the $qq$ channel, and the LHC sensitivity extends to several
TeV.  The simplest decay of such a resonance is back to two jets, and for a
diquark at 2\,TeV (or above), its contribution to the two-jet rate is comparable
with the QCD background.  The flavor physics constraints, which could be severe,
can be satisfied by making the diquark models minimally flavor
violating~\cite{Arnold:2009ay}.  As for a $Z'$, if we introduce additional new
particles, more spectacular signals are possible.  Models can be constructed in
which the final state is two new charged particles, or $2j +
\ell^+\ell^-$~\cite{Bauer:2009cc}.  The latter final state is well-studied for
$W_R^\pm$ searches, however, that has discovery potential only with $>\!1\,\fbm$
data, whereas the same $2j + \ell^+\ell^-$ search for a diquark is already
interesting with $>10\,\pbm$.

\section{Conclusions}

We explored new physics ``supermodel" scenarios that the LHC can discover with
${\cal O}(10\,\pbm)$ data.  With 1\,--\,10\,\pbm, $s$-channel resonances coupled
to $qq$ initial states (``diquarks") are the most promising, while with more
data $q\bar q$ resonance ($Z'$) searches also become interesting, especially
decaying into new charged particles.  While some supermodels may not be as
attractive as the name suggests, the same final states are useful search
channels for more conventional models later on.

\bigskip\medskip\noindent{\bf Acknowledgments}~~~
I thank C.~Bauer, M.~Schmaltz, J.~Thaler, and D.~Walker for a most enjoyable
collaboration.
This work was supported in part by the Director, Office of Science, Office
of High Energy Physics of the U.S.\ Department of Energy under contract
DE-AC02-05CH11231.

\end{document}